\documentclass[a4paper,12pt, abstract=true]{scrartcl}

\usepackage{german}
\usepackage[ansinew]{inputenc}
\usepackage[german,english,french,USenglish]{babel}
\usepackage{latexsym,amssymb,amsfonts,bbm,amsmath} 
\usepackage{theorem}
\usepackage{xcolor}
\usepackage{graphicx}
\usepackage{esint}
\usepackage{stmaryrd}
\usepackage{url}
\usepackage[normalem]{ulem} 
\newtheorem{dummytheorem}{Dummy-Theorem}[section]
\newcommand{\proofendsign}{$\Box$} 

\newtheorem{theorem}[dummytheorem]{Theorem}

\newenvironment{proof}{{\noindent \bf Proof }}
 {{\hspace*{\fill}\proofendsign\par\bigskip}}

\theorembodyfont{\normalfont}
\newtheorem{remarknorm}[dummytheorem]{Remark}
\newtheorem{examplenorm}[dummytheorem]{Example}

\newcommand{\R}{\mathbb{R}}

\newcommand{\E}{\mathbb{E}}

\newcommand{\vari}{\mathbb{V}{\rm ar}}
\newcommand{\covi}{\mathbb{C}{\rm ov}}

\newcommand{\eins}{\mathbbm{1}}

\DeclareMathOperator{\sign}{sgn}

\usepackage[comma]{natbib}

\newcount\Comments  
\newcommand{\kibitz}[2]{\ifnum\Comments=1\textcolor{#1}{#2}\fi}

\begin{document}


\title{Identifiability issues of age-period and age-period-cohort models of the Lee-Carter type}

\author{Eric Beutner$^{a}$ \and Simon Reese$^b$ \and Jean-Pierre Urbain$^c$}

\date{\footnotesize $^a$Department of Quantitative Economics,
Maastricht University, P.O.~Box 616, NL-6200 MD Maastricht,
Netherlands, {e.beutner@maastrichtuniversity.nl}\\ $^b$Department of Economics
Lund University, P.O.~Box 7082, SE-22007, Lund, Sweden, {simon.reese@nek.lu.se}\\
$^c$ Department of Quantitative Economics,
Maastricht University, P.O.~Box 616, NL-6200 MD Maastricht,
Netherlands,  {j.urbain@maastrichtuniversity.nl}\\
}
\maketitle

\begin{abstract}
The predominant way of modelling mortality rates is the Lee-Carter model and its many extensions. The Lee-Carter model and its many extensions use a latent process to forecast. These models are estimated using a two-step procedure that causes an inconsistent view on the latent variable. This paper considers identifiability issues of these models from a perspective that acknowledges the latent variable as a stochastic process from the beginning. We call this perspective the plug-in age-period or plug-in age-period-cohort model. Defining a parameter vector that includes the underlying parameters of this process rather than its realisations, we investigate whether the expected values and covariances of the plug-in Lee-Carter models are identifiable. It will be seen, for example, that even if in both steps of the estimation procedure we have identifiability in a certain sense it does not necessarily carry over to the plug-in models.
\end{abstract}

{\bf Keywords:} Time series model; Identifiability; Lee-Carter model; Plug-in Lee-Carter model; Age-period model; Age-period-cohort model.

\newpage

\section{Introduction}\label{section introduction}
Interest in the age-at-death distribution can be traced back to the works of John Graunt and Edmond Halley in 1662 and 1693, respectively; see \citep[Chapters 7 and 9]{Hald2003}.
Deriving analytical expressions for the age-at-death distribution or, which is the same, for the force of mortality goes back to the work of Gompertz in 1825 (or even to de Moivre who assumed a constant force of mortality for his work on annuities).
For further analytical expressions for the force of mortality see, for instance, \citep[Section 3.7]{Bowersetal}.
Continuing decrease of mortality rates (and consequently continuing increase of life expectancies) in many developed countries over the last six or seven decades has brought the need of forecasting mortality rates to a leading edge. A prerequisite for extrapolative methods to forecast mortality rates is a model that captures the main features of observed mortality rates. The dominant model of this approach is the Lee-Carter model (cf.~\cite{LeeCarter1992}) and its many variants; for overviews on the original model and on the many extensions that have been proposed one may refer to \cite{Booth2006}, \cite{BoothTickel2008}, \cite{Crainsetal2008}, \cite{Crainsetal2009}, \cite{Currie2014}, \cite{HabermanRenshaw2008} and \cite{HabermanRenshaw2011}, and the references therein.\\
The basic Lee-Carter model is an age-period model that takes as its starting point a non-linear parametrisation of the logarithm of the central forces of mortality. It is given by
\begin{equation}\label{eq Lee-Carter model}
\log(m_{x,t})=\alpha_x+\beta_x \kappa_t + \epsilon_{x,t},\,
x=0,\ldots,X, t=1,\ldots,T;
\end{equation}
cf.~first displayed equation in \cite[Section 3]{LeeCarter1992}. Here $m_{x,t}$ are the 'observed' central forces of mortality and $X$ is the maximal age (either in the sample or the maximum age of interest). The errors $\epsilon_{x,t}$ are assumed to have mean zero and variance $\sigma_{\epsilon}^2$. The $(X+1)$-dimensional parameter vectors $\boldsymbol{\alpha}=(\alpha_0,\ldots,\alpha_X)$ and $\boldsymbol{\beta}=(\beta_0,\ldots,\beta_X)$ are interpreted as age-specific constants.
Because in a first step $\boldsymbol{\kappa}=(\kappa_1,\ldots,\kappa_T)$ is considered to be a $T$-dimensional parameter vector, the model for the expected values of $\log(m_{x,t})$ defined by (\ref{eq Lee-Carter model}) is clearly over-parametrized.
The solution proposed by Lee and Carter to ensure identifiability of the first moments is to impose the constraints $\sum_{x=0}^X \beta_x=1$ and $\sum_{t=1}^T \kappa_t=0$, cf.~first paragraph of Section 3 in \cite{LeeCarter1992}. Under these constraints, called 'ad hoc identification' by \cite{NielsenNielsen2014}, $\alpha_x+\beta_x \kappa_t=\tilde{\alpha}_x+\tilde{\beta}_x \tilde{\kappa}_t$, $x=0,\ldots,X,t=1,\ldots,T$ imply that $\alpha_x=\tilde{\alpha}_x$, $\beta_x=\tilde{\beta}_x$, $x=0,\ldots,X$, and $\kappa_t=\tilde{\kappa}_t$, $t=1,\ldots,T$. The cohort extension of the age-period model given by equation \eqref{eq Lee-Carter model} is defined by
\begin{equation}\label{eq:LCcohort}
	\log(m_{x,t}) =  \alpha_x + \beta_x^{(0)} \iota_{t-x} + \beta_x^{(1)} \kappa_{t} + \epsilon_{x,t}.
\end{equation}
Here $\boldsymbol{\iota}=(\iota_{1-X},\ldots,\iota_T)$ represents cohort effects. This cohort extension was introduced by \cite{RenshawHaberman2006}. Again this model is over-parametrized; see section \ref{section identifiability Lee-Carter multiple factors} for more details on that.

To be able to forecast with models as given by equations (\ref{eq Lee-Carter model}) and (\ref{eq:LCcohort}) a two step procedure is applied. In a first step the parameters $\boldsymbol{\alpha}, \boldsymbol{\beta}$ and $\boldsymbol{\kappa}$ (model (\ref{eq Lee-Carter model})) or $\boldsymbol{\alpha}, \boldsymbol{\beta}^{(0)}, \boldsymbol{\beta}^{(1)}, \boldsymbol{\kappa}$ and $\boldsymbol{\iota}$ (model (\ref{eq:LCcohort})), respectively, are estimated. In the second step a time series model that allows for forecasting is fitted to the estimated $\hat{\boldsymbol{\kappa}}$-vector or to the estimated $\hat{\boldsymbol{\kappa}}$ and $\hat{\boldsymbol{\iota}}$-vectors. In the following, we will refer to the classical Lee-Carter model and its cohort extension as fully parametric age-period or age-period-cohort Lee-Carter models, respectively. This denomination relates to the treatment of estimated age and cohort effects as factor scores, i.e. estimates of $T$ and $T+X$ dimensional parameter vectors, in step one of the estimation process (see e.g. \citet[Section 13.6]{Rencher2002} for the notion of factor scores).  Imposing a stochastic model on the estimates of these parameters in step two is conceptually inconsistent and leads to problems when specifying identifiability constraints. For example, it is a priori unclear whether the forecast from the imposed stochastic model depends on the chosen identification scheme for the original parameters $\boldsymbol{\alpha}, \boldsymbol{\beta}$ and $\boldsymbol{\kappa}$ (model (\ref{eq Lee-Carter model})) or $\boldsymbol{\alpha}, \boldsymbol{\beta}^{(0)}, \boldsymbol{\beta}^{(1)}, \boldsymbol{\kappa}$ and $\boldsymbol{\iota}$ (model (\ref{eq:LCcohort})). For model (\ref{eq Lee-Carter model}) this question has been addressed by \cite{NielsenNielsen2014} who build on \cite{Kuangetal2008b} where the same question is analysed for an additive age-period-cohort model. Moreover, as detailed in section 2, the 'ad hoc identification' constraints lead to implausible constraints on the properties of the stochastic model imposed onto the factor scores.
An alternative perspective on the Lee-Carter model is to replace $\boldsymbol{\kappa}$, or $\boldsymbol{\kappa}$ and $\boldsymbol{\iota}$, by time series models from the beginning. We will denote these models as plug-in age-period and age-period-cohort Lee-Carter models respectively. Plug-in age-period Lee-Carter models have so far been considered in \cite{GirosiKing2007} and \cite{DeJongTickle2006}.

Recently, \cite{LengPeng2016} considered a simplified fully parametric age-period Lee-Carter model and showed that the two step estimation procedure may lead to inconsistent estimators. A pre-requisite for consistency is identifiability. This paper, therefore, considers the interplay between identifiability of fully parametric and plug-in Lee-Carter models. Suppose that we have an identification scheme for a fully parametric Lee-Carter model. Furthermore, suppose that we use an identifiable time series model for $\boldsymbol{\kappa}$ or identifiable time series models for $\boldsymbol{\kappa}$ and $\boldsymbol{\iota}$, respectively. Do the plug-in Lee-Carter models inherit identifiability from identifiability of the fully parametric Lee-Carter models and the identified time series models? We show that this needs not be the case.
Furthermore, assume that identifiable time series models are plugged in into a non-identified fully parametric Lee-Carter model. Is it possible that the resulting plug-in Lee-Carter model is nevertheless identified? We will see that this possibility can occur. We address these two questions in section \ref{section identifiability Lee-Carter random walk model} by considering simple but very popular times series models for $\boldsymbol{\kappa}$ and $\boldsymbol{\iota}$. 
More precisely, we first look at the age-period model \eqref{eq Lee-Carter model} if a random walk is used to model the factor scores for $\boldsymbol{\kappa}$. Afterwards we analyse the age-period-cohort Lee-Carter model (\ref{eq:LCcohort}) if two independent random walks are used to model the factor scores for $\boldsymbol{\kappa}$ and $\boldsymbol{\iota}$. Having addressed the above questions and having obtained identifiability results if random walks are plugged in we briefly extend our considerations to more complicated time series models because from an applied point it is important that the class of time series models for which  plug-in models are identifiable is not too narrow. This will be done in section \ref{section extensions}. Our findings concerning identifiability of plug-in models continue to hold for generalized linear models whose index function is modelled in the fashion of equations \eqref{eq Lee-Carter model} and \eqref{eq:LCcohort}.

The rest of the article is organised as follows:  In section \ref{subsection discussion constraints} we briefly discuss inconsistencies that arise from the two step procedure. 
Section \ref{section identifiability Lee-Carter random walk model} and \ref{section extensions} are as described above. We conclude with a discussion. All proofs are presented in the appendix.

\section{Stochastic process view on the Lee-Carter constraints}\label{subsection discussion constraints}
If we impose a stochastic model on $(\kappa_t)$ as done in the statistical analysis of the fully parametric Lee-Carter model, the identifying restriction $\sum_{t=1}^T \kappa_t=0$ becomes a constraint on the possible realizations of the stochastic process $(\kappa_t)$. As such the constraint does not seem to be sensitive, because it implies inconsistencies in the modelling procedure. An early reference that differentiates sets of constraints depending on whether the factor(s) is\slash are assumed to be an unobserved random process or unobserved but deterministic is \cite{AndersenRubin1956}. Here we examine two inconsistencies that arise if we constrain the realizations of the stochastic process $(\kappa_t)$.
\begin{enumerate}
\item {\it Dynamic view on the constraint:}  Suppose that we estimated the model based on data up to and including $\breve{T}$ and that we now want to update our estimates based on data up to and including $\breve{T}+1$. If the realization of $(\kappa_1,\ldots,\kappa_{\breve{T}})$ fulfils the constraint, then we must have $\kappa_{\breve{T}+1}=0$, because otherwise $\sum_{t=1}^{\breve{T}+1} \kappa_{t}=0$ is impossible. This is because we cannot change the realization of $(\kappa_1,\ldots,\kappa_{\breve{T}})$ which is given to us. This is different from increasing $X$, because $\boldsymbol{\beta}$ is part of the modelling process and not given exogenously to us as the realization of a stochastic process. Notice also that the same reasoning applied sequentially to $\breve{T}+(k-1)$, $k \geq 2$, would imply $\kappa_{\breve{T}+k}=0$, $k \geq 2$.

\item {\it Distributional view on the constraint:} Assume, for instance, that the outcome of the second step of the statistical analysis done by Lee and Carter is that $(\kappa_t)$ follows a random walk with or without drift. Assume additionally that the random walk starts in $c \in \R$, i.e.~$\kappa_0=c$, and that the innovations are normally distributed or more general that the joint distributions of the innovations possess a probability density function with respect to Lebesgue measure then the event
$
\{\sum_{t=1}^T \kappa_t=0\}
$
has probability zero for every $T \geq 1$ regardless of the starting value $c$, because $\{x \in \R^{T}| \sum_{i=1}^{T} x_i=0\}$ is a hyperplane in $\R^{T}$. Consequently, under these assumptions the probability that the constraint is fulfilled equals zero.
\end{enumerate}


\section{Identifiability of plug-in age-period and age-period-cohort Lee-Carter models}\label{section identifiability Lee-Carter random walk model}
\subsection{Preliminaries}\label{subsection preliminaries}
Throughout and irrespective of whether we consider an age-period or an age-period cohort model we assume that
\begin{align*}\label{eq definition identifiability}
& \E_{\boldsymbol{\theta}}(\log(m_{x,t}))=f_{\boldsymbol{\theta}}(x,t), x=0,\ldots,X, t=1,\ldots,T, \nonumber \\
&\covi_{\boldsymbol{\theta}}(\log(m_{x,s}),\log(m_{y,t}))=g_{\boldsymbol{\theta}}(x,y,s,t), x,y \in \{0,\ldots,X\}, s,t \in \{1,\ldots,T\},
\end{align*}
for known functions $f$ and $g$, and some unknown parameter $\boldsymbol{\theta} \in \Theta$ with the parameter space $\Theta$ being a subset of some finite-dimensional space. For some models considered below $f_{\boldsymbol{\theta}}$ depends only on $t$ and not on $x$. However, in the more general case $f_{\boldsymbol{\theta}}$ depends on $x$ and $t$. Similar for $g_{\boldsymbol{\theta}}$. We say that the expected values are identifiable if $f_{\boldsymbol{\theta}}(x,t)=f_{\tilde{\boldsymbol{\theta}}}(x,t)$ for $x=0,\ldots,X$, $t=1,\ldots,T$ implies: $\boldsymbol{\theta}=\tilde{\boldsymbol{\theta}}$. Similarly, we say that the expected values and the covariance structure are identifiable if $f_{\boldsymbol{\theta}}(x,t)=f_{\tilde{\boldsymbol{\theta}}}(x,t)$ for $x=0,\ldots,X$, $t=1,\ldots,T$ and $g_{\boldsymbol{\theta}}(x,y,s,t)=g_{\tilde{\boldsymbol{\theta}}}(x,y,s,t)$ for $x,y \in \{0,\ldots,X\}$, $s,t \in \{1,\ldots,T\},$   together imply:  $\boldsymbol{\theta}=\tilde{\boldsymbol{\theta}}$.
\subsection{Identifiability of plug-in age-period models}
In this section we analyse plug-in age-period Lee-Carter models
if the plug-in process is a random walk, i.e.~we assume
$\kappa_t=\mu+\kappa_{t-1}+e_t=\mu\,t+c+\sum_{\ell=1}^t e_{\ell}$, $t \geq 1$, with $\kappa_0=c$, $c \in \R$, and  $(e_t)$ is a sequence of independent and identically distributed random variables with $\E(e_t)=0$ that is independent of $\epsilon_{x,t}$, $x=0,\ldots,X,t=1,\ldots,T$. Clearly, there is a one-to-one correspondence between $\mu \in \R$ and $\E(\kappa_t)=c+\mu t$ so that the expected values are identifiable. Looking at the plug-in age-period Lee-Carter model with $\boldsymbol{\kappa}$ being a random walk we have
\begin{equation}\label{eq central death rates with random walk}
\log(m_{x,t})=\alpha_x+\beta_x \mu \,t+\beta_x c + \beta_x \sum_{\ell=1}^t e_{\ell} + \epsilon_{x,t},\quad x=0,\ldots,X,t=1,\ldots,T.
\end{equation}
Clearly there is no one-to-one correspondence between $\{(\boldsymbol{\alpha},\boldsymbol{\beta},\mu) \in \R^{X+1} \times \R^{X+1} \times \R| \sum_{x=0}^X \beta_x=1\}$ and the expected values of $\log(m_{x,t})$. A counterexample is given by $\mu=0$ and any two vectors $\boldsymbol{\beta} \neq \tilde{\boldsymbol{\beta}}$ that both fulfil the identification constraint if we put $\tilde{\boldsymbol{\alpha}}=(\alpha_0+c(\beta_0-\tilde{\beta}_0),\ldots,\alpha_X+c(\beta_X-\tilde{\beta}_X))$. This reveals that identifiable expected values of a fully parametric age-period model combined with a time series model with identifiable expected values do not lead to identifiable expected values for the plug-in age-period model. Here, the problem can be overcome by excluding $\mu=0$ or by taking other moments into account. Assume that $\E(e_t^2)=\sigma_e^2$ and consider (\ref{eq central death rates with random walk}) to be a model with parameter set
$$\Theta_{(0,1,0)}:=\left\{(\boldsymbol{\alpha},\boldsymbol{\beta},\mu,\sigma_{e}^2, \sigma_{\epsilon}^2) \in \R^{(X+1)} \times \R^{(X+1)} \times \R 
\times \R_+ \times \R_+| \sum_{x=0}^X \beta_x=1\right\},$$
where $\R_+=\{x \in \R| x>0\}$. For all $\boldsymbol{\theta} \in \Theta$ the expected values and the covariances of (\ref{eq central death rates with random walk}) are given by
\begin{equation}\label{eq identifiability proof means}
		\E_{\boldsymbol{\theta}}(\log(m_{x,t}))=\alpha_x+\beta_x \mu \,t+\beta_x c,\quad x=0,\ldots,X,t=1,\ldots,T,
	\end{equation}
	and
	\begin{align}
	\label{eq identifiability proof covariances}
		\covi_{\boldsymbol{\theta}}(\log(m_{x,s}),\log(m_{y,t}))=\beta_x \beta_y \sigma_{e}^2\,\min\{s,t\}+ \eins_{\{x=y, s=t\}}(x,y,s,t)\sigma_{\epsilon}^2, \\
		x, y \in \{0,\ldots,X\},\,s,t \in \{1,\ldots,T\}, \notag
	\end{align}
respectively, and we have the following result
\begin{theorem} \label{thm:LC_rw}
The expected values and the covariance structure of the model given by
	(\ref{eq central death rates with random walk}) with parameter set $\Theta_{(0,1,0)}$ are identifiable in the sense of section \ref{subsection preliminaries} if $T \geq 2$.
\end{theorem}

\subsection{Identifiability of plug-in age-period-cohort models}\label{section identifiability Lee-Carter multiple factors}

Identifiability of fully parametric age-period-cohort Lee-Carter models has recently been questioned by \cite{HuntVillegas2015} who refer to convergence problems of the maximum likelihood estimator and lack of robustness of the estimates in a number of studies. The identifying restrictions used in these models are usually chosen to be
$$\sum_{x=0}^X \beta_x^{(0)}=\sum_{x=0}^X \beta_x^{(1)}=1,\, \kappa_{1}=0,\, \beta_x^{(1)}>0,\, x=0,1,\ldots,X$$
or
$$\sum_{x=0}^X \beta_x^{(1)}=\sum_{x=0}^X \beta_x^{(0)}=1,\, \sum_{t=1}^T \kappa_t=0,\,\sum_{h=-X+1}^T \iota_h=0;$$
see, for instance, \citep[equation (2)]{HabermanRenshaw2009} and \citep[Section 3]{Yangetal2014}.
In fact, as example
\ref{example cohort fitting constraints} below shows, 
 none of the two sets of constraints above ensures that $(\boldsymbol{\alpha},\boldsymbol{\beta}^{(1)},\boldsymbol{\beta}^{(0)},\boldsymbol{\kappa},\boldsymbol{\iota}) \neq
(\tilde{\boldsymbol{\alpha}},\tilde{\boldsymbol{\beta}}^{(1)},\tilde{\boldsymbol{\beta}}^{(0)},\tilde{\boldsymbol{\kappa}},
\tilde{\boldsymbol{\iota}})$ implies
$\alpha_x+\beta_x^{(1)}\kappa_{t}+\beta_x^{(0)}\iota_{t-x} \neq \tilde{\alpha}_x+\tilde{\beta}_x^{(1)}\tilde{\kappa}_{t}+\tilde{\beta}_x^{(0)}\tilde{\iota}_{t-x} \mbox{ for at least one pair } (x,t).$

\begin{examplenorm}\label{example cohort fitting constraints}
Let $X>2,T>2$. Take $(\boldsymbol{\alpha},\boldsymbol{\beta}^{(1)},\boldsymbol{\kappa})$ such that the constraints on $\boldsymbol{\beta}^{(1)}$ and $\boldsymbol{\kappa}$ are fulfilled. Put
$
(\tilde{\boldsymbol{\alpha}},\tilde{\boldsymbol{\beta}}^{(1)},\tilde{\boldsymbol{\kappa}})=(\boldsymbol{\alpha},\boldsymbol{\beta}^{(1)},\boldsymbol{\kappa}).
$
Next define $\boldsymbol{\beta}^{(0)}=(\beta_0^{(0)},\ldots,\beta_X^{(0)})$ and $\boldsymbol{\iota}=(\iota_{-X+1},\ldots,\iota_T)$ by
\begin{align*}
& \beta_0^{(0)}=0.75,\beta_1^{(0)}=0.25, \beta_x^{(0)}=0, x=2,\ldots,X; \\
& \iota_{h}=0 \mbox{ for }  h=-X+2,\ldots,-1,1,\ldots,T-1,\,\iota_{-X+1}=-2,\iota_0=1,\iota_T=1.
\end{align*}
Then the constraints are clearly fulfilled and
\begin{equation*}
\beta_x^{(0)} \iota_{t-x} = \left\{
\begin{array}{cc}
0.25, & x=1,t=1\\
0.75, & x=0,t=T\\
0, & \mbox{otherwise.} 
\end{array}
\right.
\end{equation*}
Finally define $\tilde{\boldsymbol{\beta}}^{(0)}=(\tilde{\beta}_0^{(0)},\ldots,\tilde{\beta}_X^{(0)})$ and $\tilde{\boldsymbol{\iota}}=(\tilde{\iota}_{-X+1},\ldots,\tilde{\iota}_T)$ by
\begin{align*}
& \tilde{\beta}_0^{(0)}=0.5,\tilde{\beta}_1^{(0)}=0.5, \tilde{\beta}_x^{(0)}=0, x=2,\ldots,X; \\
& \tilde{\iota}_{h}=0 \mbox{ for }  h=-X+2,\ldots,-1,1,\ldots,T-1,\,\tilde{\iota}_{-X+1}=-2,\tilde{\iota}_0=0.5,\tilde{\iota}_T=1.5.
\end{align*}
 Again the constraints are fulfilled and $\tilde{\beta}_x^{(0)} \tilde{\iota}_{t-x} = \beta_x^{(0)} \iota_{t-x}$ for all $x=0,1,\ldots,X,\,t=1,\ldots,T$. Apparently, $(\tilde{\boldsymbol{\beta}}^{(0)}, \tilde{\boldsymbol{\iota}}) \neq (\boldsymbol{\beta}^{(0)},\boldsymbol{\iota})$ and therefore
$$(\boldsymbol{\alpha},\boldsymbol{\beta}^{(1)},\boldsymbol{\beta}^{(0)},\boldsymbol{\kappa},\boldsymbol{\iota}) \neq
(\tilde{\boldsymbol{\alpha}},\tilde{\boldsymbol{\beta}}^{(1)},\tilde{\boldsymbol{\beta}}^{(0)},\tilde{\boldsymbol{\kappa}},
\tilde{\boldsymbol{\iota}}),$$
but by construction $\alpha_x+\beta_x^{(1)}\kappa_{t}+\beta_x^{(0)}\iota_{t-x} = \tilde{\alpha}_x+\tilde{\beta}_x^{(1)}\tilde{\kappa}_{t}+\tilde{\beta}_x^{(0)}\tilde{\iota}_{t-x}$ for all $x=0,1,\ldots,X,\,t=1,2,\ldots,T$.
\end{examplenorm}

What does the non-identifiability of fully parametric age-period-cohort Lee-Carter models imply for their plug-in counterpart? It turns out that the plug-in model actually is identifiable. Let $\kappa_t$ and $\iota_{t-x}$ be given by

\begin{align}
	\label{eq:LCcohortRW}
	\kappa_t = \kappa_0 + t\mu_1 + \sum_{s=1}^t e^{(1)}_s,\,\quad
	\iota_{t-x} = \iota_{-X} + (t-x+X)\mu_0 + \sum_{r = 1}^{t-x+X} e^{(0)}_{r-X}
\end{align}
where $(e^{(1)}_s)$ and $(e^{(0)}_t)$ are two independent sequences of independent and identically distributed random variables with expected values equal to zero and finite second moments denoted by $\sigma_{e_1}^2$ and $\sigma_{e_2}^2$, respectively; see, for example, \citep[Section 6.2]{HabermanRenshaw2008}. The model that results if $(\kappa_1,\ldots,\kappa_T)$ and $(\iota_{-X+1},\ldots,\iota_T)$ in equation (\ref{eq:LCcohort}) are assumed to be given by equation (\ref{eq:LCcohortRW}) is considered to be a model with parameter set
\begin{align*}
\Theta_{(0,1,0)\times(0,1,0)}^{ct}  := &  \Big\{(\boldsymbol{\alpha},\boldsymbol{\beta}^{(0)},\boldsymbol{\beta}^{(1)},\mu_0,\mu_1,\sigma_{e_1}^2,\sigma_{e_2}^2, \sigma_{\epsilon}^2)
\\ & \hspace{0.5cm}
  \in \R^{X+1} \times \R^{X+1} \times \R^{X+1} \times \R \times \R  \times \R_+ \times \R_+ \times \R_+
  \\ &    \hspace{0.5cm}
   | \sum_{x=0}^X \beta_x^{(i)}=1,\,i=0,1,\,\boldsymbol{\beta}^{(0)} \neq \boldsymbol{\beta}^{(1)} \big\}.
\end{align*}
With $\iota_{-X} = c_0$ and $\kappa_0 = c_1$ the expected values of the model defined by \eqref{eq:LCcohort} and \eqref{eq:LCcohortRW} are given by
\begin{equation}\label{eq expected values cohort random walk}
	\E_{\boldsymbol{\theta}}(\log(m_{x,t}))=\alpha_x + \beta_x^{(0)}c_0 + \beta_x^{(0)}\mu_0 (t-x+X)  + \beta_x^{(1)}c_1 + \beta_x^{(1)}\mu_1 t,
\end{equation}
and the covariance structure by
\begin{align}\label{eq covariance cohort plug-in}
	\covi_{\boldsymbol{\theta}}(\log(m_{x,s}), \log(m_{y,t}))	= & \beta_x^{(0)} \beta_y^{(0)} \sigma_{e_0}^2\,\min\{s-x+X,t-y+X\}+\beta_x^{(1)} \beta_y^{(1)} \sigma_{e_1}^2\,\min\{s,t\} \notag \\
	& + \eins_{\{x=y, s=t\}}(x,y,s,t)\sigma_{\epsilon}^2.
\end{align}
We have the following result:


\begin{theorem}\label{thm:LC_rw 2}
   The expected values and the covariance structure of the model defined by \eqref{eq:LCcohort} and \eqref{eq:LCcohortRW} with parameter set $\Theta_{(0,1,0) \times (0,1,0)}^{ct}$ are identifiable in sense of section \ref{subsection preliminaries} if $T>X+2$ and $X>0$.
\end{theorem}

\begin{remarknorm}\label{remark cohort}
(i) Without $\boldsymbol{\beta}^{(0)} \neq \boldsymbol{\beta}^{(1)}$ identifiability fails. Let $\mu_0 \neq \mu_1$, take 
 $\tilde{\boldsymbol{\beta}}^{(0)}=\tilde{\boldsymbol{\beta}}^{(1)}=\boldsymbol{\beta}^{(0)}=\boldsymbol{\beta}^{(1)}, \tilde{\mu}_0=\mu_1, \tilde{\mu}_1=\mu_0, \tilde{\sigma}_{e_1}^2=\sigma_{e_1}^2,\tilde{\sigma}_{e_2}^2=\sigma_{e_2}^2, \tilde{\sigma}_{\epsilon}^2=\sigma_{\epsilon}^2$, and put $\tilde{\alpha}_x=\alpha_x-(X-x)\tilde{\beta}_x^{(0)} \tilde{\mu}_0+(X-x)\tilde{\beta}_x^{(1)}\mu_0$.\\
 (ii) Furthermore, $X>0$ cannot be dropped. Indeed  if $X=0$ the covariance structure (cf.~equation (\ref{eq covariance cohort plug-in})) reduces to
$\covi(\log(m_{0,s}), \log(m_{0,t})) =  (\sigma_{e_0}^2 + \sigma_{e_1}^2)\,\min\{s,t\} + \sigma_{\epsilon}^2$. Here, $\{\tilde{\sigma}_{e_0}^2,  \tilde{\sigma}_{e_1}^2,  \tilde{\sigma}_{\epsilon}^2 \}$ with $\tilde{\sigma}_{\epsilon}^2 = \sigma_{\epsilon}^2$, $\tilde{\sigma}_{e_0}^2 = \sigma_{e_0}^2 + z$ and $\tilde{\sigma}_{e_1}^2 = \sigma_{e_1}^2 - z$ for any $z \in (-\sigma_{e_0}^2,  \sigma_{e_1}^2)$ is a valid reparameterisation.\\ 
(iii) In contrast to this we do not know whether the constraint $T>X+2$ that is employed in the proof of theorem \ref{thm:LC_rw 2} is necessary.
\end{remarknorm}

\section{Some extensions}\label{section extensions}
Even though mortality rates are predominantly forecasted by fitting a simple random walk model on the estimated latent
trend, 
some studies use other simple autoregressive integrated moving average processes for this purpose. Examples are the autoregressive integrated moving average processes with parameter (1,1,0) (ARIMA(1,1,0)) applied in \cite{Cairnsetal2005} and the autoregressive integrated moving average processes with parameter (0,1,1) (ARIMA(0,1,1)) in \cite{Brouhnsetal2005} as well as \cite{CoelhoNunes2011} who use both models. Under ARIMA(1,1,0) we have $(1-\rho L)(\Delta\kappa -\mu) = e_t$. Here $L$ is the lag operator. In this case, as detailed in appendix \ref{appendix ARIMA(1,1,0)} we have that equation (\ref{eq Lee-Carter model}) becomes
\begin{equation}
	\label{eq:LCARIMA110}
	\log(m_{x,t})=\alpha_x+\beta_x \mu \,t+\beta_x c + \beta_x \sum_{\ell=1}^t \sum_{k=0}^{\infty} \rho^k e_{\ell-k} + \epsilon_{x,t},\quad x=0,\ldots,X,t=1,\ldots,T.
\end{equation}
We consider \eqref{eq:LCARIMA110} to be a model with parameter set
\begin{align*}
\Theta_{(1,1,0)}:=\Big\{(& \boldsymbol{\alpha},\boldsymbol{\beta},\mu, \rho, \sigma_{e}^2, \sigma_{\epsilon}^2) \in \R^{X+1} \times \R^{X+1} \times \R \times (-1,1) \times \R_+ \times \R_+
| \sum_{x=0}^X \beta_x=1\Big\}.
\end{align*}
We have the following result similar to theorem \ref{thm:LC_rw}
\begin{theorem}\label{prop:LC_ARIMA110}
	The expected values and the covariance structure of the model given by \eqref{eq:LCARIMA110} with parameter set $\Theta_{(1,1,0)}$ are identifiable in the sense of section \ref{subsection preliminaries} if $T \geq 4$.
\end{theorem}

Lastly, assuming that the latent factor follows an ARIMA(0,1,1) process amounts to letting $ \Delta\kappa_t - \mu = e_t + \phi e_{t-1}$. Consequently, \eqref{eq Lee-Carter model} becomes
\begin{equation}
	\label{eq:LCARIMA011}
	\log(m_{x,t})=\alpha_x+\beta_x \mu \,t+\beta_x c + \beta_x \sum_{\ell=1}^t (e_{\ell} +  \phi e_{\ell-1}) +  \epsilon_{x,t},\quad x=0,\ldots,X,t=1,\ldots,T.
 \end{equation}
 We consider \eqref{eq:LCARIMA011} to be a model with parameter set
\begin{align*}
\Theta_{(0,1,1)}:=\Big\{(& \boldsymbol{\alpha},\boldsymbol{\beta},\mu, \phi, \sigma_{e}^2, \sigma_{\epsilon}^2) \in \R^{X+1} \times \R^{X+1} \times \R \times (-1,1) \times \R_+ \times \R_+ | \sum_{x=0}^X \beta_x=1\Big\}.
\end{align*}

We have the following result:

\begin{theorem}\label{prop:LC_ARIMA011}
The expected values and the covariance structure of the model given by \eqref{eq:LCARIMA011} with parameter set $\Theta_{(0,1,1)}$ are identifiable in the sense of section \ref{subsection preliminaries} if $T \geq 2$.
\end{theorem}

\section{Discussion}	\label{section discussion}
Extensions of the seminal model of Lee and Carter (1992) are by now so plentiful that they constitute an own branch of the literature on mortality modelling. However, despite the large number of contributions that focus on improving the performance of the standard model, research on the properties of the pervasive features of all these models is still scarce. This article considers identifiability of age-period and age period-cohort Lee-Carter models and their extensions from a perspective that directly acknowledges the stochastic nature of the latent trend in mortality. The relevance of viewing models from this perspective, resulting in what we call the plug-in Lee Carter model, is emphasized by logical inconsistencies that arise from the standard practice of treating the latent trend in mortality as a parameter vector. We find that the predominant set of identifying restrictions need not ensure identifiability of plug-in age-period Lee Carter models. On the contrary, age-period-cohort plug-in Lee Carter models may be identified under restrictions that are insufficient to ensure identifiability of their equivalent under the classical perspective. Still, despite the findings provided in this paper, much works remains to be done on the theoretical properties of the Lee-Carter model and its extensions. Investigating these is of utmost importance for further contributions on estimation of the model and the conclusions that can be drawn from its parameter values.


\appendix

\section{Appendix}
This appendix contains the proofs of the theorems given above. Regarding the proofs for age-period models it is worth mentioning that a technique different from ours could be used: First demean these models, second apply results from \cite{HeatonSolo2004} to identify the covariance structure up to some indeterminacies, third use the results of the second step and the constraints on $\boldsymbol{\beta}$ to identify the expected values and to overcome the mentioned indeterminacies. We do not follow this approach as it cannot be applied to age-period-cohort models and it would not simplify the proofs. Moreover, \cite{HeatonSolo2004} consider the covariance structure to be identifiable if $g_{\boldsymbol{\theta}}(x,y,s,t)=g_{\tilde{\boldsymbol{\theta}}}(x,y,s,t)$, $x,y \in \{0,\ldots,X\}$,$s,t \in \{1,2,\ldots\}$ implies $\boldsymbol{\theta}=\tilde{\boldsymbol{\theta}}$ which does not allow to directly determine a finite $T$ as it is done in theorems \ref{thm:LC_rw}, \ref{prop:LC_ARIMA110} and \ref{prop:LC_ARIMA011}. \\

\subsection{Proofs of theorems  \ref{thm:LC_rw} and \ref{thm:LC_rw 2}}

\begin{proof} {\bf of theorem  \ref{thm:LC_rw}}
Let $\boldsymbol{\theta}, \boldsymbol{\theta} \in \Theta$	and assume $f_{\boldsymbol{\theta}}(x,t)=f_{\tilde{\boldsymbol{\theta}}}(x,t)$ and $g_{\boldsymbol{\theta}}(x,y,s,t)=g_{\tilde{\boldsymbol{\theta}}}(x,y,s,t)$. For $x=y$ and $t=s$ using expression (\ref{eq identifiability proof covariances}) the latter implies $\beta^2_x \sigma^2_e\, t + \sigma^2_{\epsilon}=\tilde{\beta}^2_x \tilde{\sigma}^2_e\, t + \tilde{\sigma}^2_{\epsilon}$. Because both are affine-linear functions and $T \geq 2$ we must have
	$\beta^2_x \sigma^2_e=\tilde{\beta}_x^2 \tilde{\sigma}^2_e$ and $\sigma^2_{\epsilon}=\tilde{\sigma}^2_{\epsilon}$.
	For $X=0$ we 
	obtain $\sigma^2_e=\tilde{\sigma}^2_e$, because then $\beta_0=\tilde{\beta}_0=1$. 
	Furthermore, for $X=0$ expression (\ref{eq identifiability proof means}) yields $(\alpha_0+c)+\mu t = (\tilde{\alpha}_0+c) +\tilde{\mu}t$ which implies $\boldsymbol{\alpha}=\tilde{\boldsymbol{\alpha}}$ and $\mu=\tilde{\mu}$, because $T \geq 2$.\\
	For $X \geq 1$, it is enough to restrict attention to those $\beta_x$ that are unequal 0, because if $\beta_x=0$ we must also have $\tilde{\beta}_x=0$, otherwise $\beta^2_x \sigma^2_e=\tilde{\beta}_x^2 \tilde{\sigma}^2_e$ is impossible since $\sigma^2_e>0$ and $\tilde{\sigma}^2_e>0$.
Put $\breve{X}:=\{x \in \{0,\ldots,X\} | \beta_x \neq 0\}$.
For $t=T$ and $x \neq y$ expression (\ref{eq identifiability proof covariances}) implies that $\beta_x \beta_y \sigma_{e}^2\, s=\tilde{\beta}_x \tilde{\beta}_y \tilde{\sigma}_{e}^2 \, s$ for  $x, y \in \breve{X}$ and $s \in \{1,\ldots,T\}$. For $x$ and $y$ fixed this is a linear function in $s$, hence:  $\beta_x \beta_y \sigma_{e}^2=\tilde{\beta}_x \tilde{\beta}_y \tilde{\sigma}_{e}^2$, $x,y \in \breve{X}$. This can only hold if either $\sign(\beta_x)=\sign(\tilde{\beta}_x)$, $\forall x \in \breve{X}$, or if $\sign(\beta_x)=-\sign(\tilde{\beta}_x)$, $\forall x \in \breve{X}$, where $\sign$ denotes the signum function. In the latter case we have $- K \tilde{\beta}_x = \beta_x$ for the constant $K = \sqrt{\tilde{\sigma}_e^2/\sigma_e^2} $, which leads to a contradiction since $1=\sum_{x=0}^X \beta_x=-K\sum_{x=0}^X \tilde{\beta}_x=-K$ is impossible as $K>0$. The same reasoning shows that if $\sign(\beta_x)=\sign(\tilde{\beta}_x)$, $\forall x \in \breve{X}$, we must have $K=1$ which implies $\beta_x=\tilde{\beta}_x$ and $\sigma_e^2=\tilde{\sigma}_e^2$. Using $\beta_x=\tilde{\beta}_x$ expression  (\ref{eq identifiability proof means}) implies $(\alpha_x+c\beta_x)+\beta_x \mu t = (\tilde{\alpha}_x+c\tilde{\beta}_x)+\tilde{\beta}_x \tilde{\mu} t$ which leads to $\alpha_x=\tilde{\alpha}_x$ and $\mu=\tilde{\mu}$ as $T \geq 2$ and $\beta_x=\tilde{\beta}_x \neq 0$ for at least one $x \in \{0,\ldots,X\}$.
\end{proof}


\begin{proof}{\bf of theorem \ref{thm:LC_rw 2}}
Let $\boldsymbol{\theta}, \tilde{\boldsymbol{\theta}} \in \Theta$	and assume $f_{\boldsymbol{\theta}}(x,t)=f_{\tilde{\boldsymbol{\theta}}}(x,t)$ and $g_{\boldsymbol{\theta}}(x,y,s,t)=g_{\tilde{\boldsymbol{\theta}}}(x,y,s,t)$. For $x=y, s=t$ expression (\ref{eq covariance cohort plug-in}) leads to
\begin{align*}
&
((\beta_x^{(0)})^2 \sigma_{e_0}^2 + (\beta_x^{(1)})^2 \sigma_{e_1}^2 )\,t -  (\beta_x^{(0)})^2 \sigma_{e_0}^2\,(x-X)  + \sigma_{\epsilon}^2
\\ &
 =((\tilde{\beta}_x^{(0)})^2 \tilde{\sigma}_{e_0}^2 + (\tilde{\beta}_x^{(1)})^2 \tilde{\sigma}_{e_1}^2 )\,t -  (\tilde{\beta}_x^{(0)})^2 \tilde{\sigma}_{e_0}^2\,(x-X)  + \tilde{\sigma}_{\epsilon}^2.
\end{align*}
Because for $x$ fixed both functions are affine linear in $t$ and by assumption $T>1$, we obtain by considering the case $x=X$ that
$\sigma_{\epsilon}^2=\tilde{\sigma}_{\epsilon}^2$. Then by the same argument we get $(\beta_x^{(0)})^2 \sigma_{e_0}^2=(\tilde{\beta}_x^{(0)})^2 \tilde{\sigma}_{e_0}^2$ for $x=0,\ldots,X-1$. As in the proof of theorem \ref{thm:LC_rw}  we obtain $|\tilde{\beta}_x^{(0)}|=c|\beta_x^{(0)}|$, $x=0,\ldots,X-1$. Now taking $x=X$, $y=X-1$ and $s=t+2$ the right-hand side of (\ref{eq covariance cohort plug-in}) equals
\begin{eqnarray}\label{eq proof cohort X and X-1}
 \beta_X^{(0)} \beta_{X-1}^{(0)} \sigma_{e_0}^2(t+1)+\beta_X^{(1)} \beta_{X-1}^{(1)} \sigma_{e_1}^2\,t
	&=& \beta_X^{(0)} \beta_{X-1}^{(0)} \sigma_{e_0}^2+(\beta_X^{(1)} \beta_{X-1}^{(1)} \sigma_{e_1}^2+\beta_X^{(0)} \beta_{X-1}^{(0)} \sigma_{e_0}^2)\,t \nonumber \\
\end{eqnarray}
Because by assumption $T \geq 4$ we have $s \leq T$ for at least $t=1$ and $t=2$ so that (\ref{eq proof cohort X and X-1}) holds for at least two different values of $t$. Then in view of $|\beta_{X-1}^{(0)} \sqrt{\sigma_{e_0}^2}|=|\tilde{\beta}_{X-1}^{(0)} \sqrt{\tilde{\sigma}_{e_0}^2}|$ and our assumption $g_{\boldsymbol{\theta}}(x,y,s,t)=g_{\tilde{\boldsymbol{\theta}}}(x,y,s,t)$ it implies $|\beta_{X}^{(0)} \sqrt{\sigma_{e_0}^2}|=|\tilde{\beta}_{X}^{(0)} \sqrt{\tilde{\sigma}_{e_0}^2}|
$.
Hence $\beta_{X}^{(0)}$ and $\tilde{\beta}_{X}^{(0)}$ differ by the same multiplicative constant $c$ as $\beta_{x}^{(0)}$ and $\tilde{\beta}_{x}^{(0)}$, $x=0,\ldots,X-1$, do.
Now for $k=1,\ldots,X$ taking $x=X$, $y=X-k$ and $s=t+k+1$ the right-hand side of (\ref{eq covariance cohort plug-in}) equals
\begin{eqnarray}\label{proof cohort eq sign}
 \beta_X^{(0)} \beta_{X-k}^{(0)} \sigma_{e_0}^2(t+k)+\beta_X^{(1)} \beta_{X-k}^{(1)} \sigma_{e_1}^2\,t
	= k \beta_X^{(0)} \beta_{X-k}^{(0)} \sigma_{e_0}^2+(\beta_X^{(1)} \beta_{X-k}^{(1)} \sigma_{e_1}^2+\beta_X^{(0)} \beta_{X-k}^{(0)} \sigma_{e_0}^2)\,t. \nonumber \\
	\end{eqnarray}
The constraint $T>X+2$ ensures that $s=t+k+1$, $k=1,\ldots,X$, takes at least two different values that are less than or equal $T$. Hence, for $X-k$ fixed (\ref{proof cohort eq sign}) holds for at least two different values of $t$. Therefore, the assumption $g_{\boldsymbol{\theta}}(x,y,s,t)=g_{\tilde{\boldsymbol{\theta}}}(x,y,s,t)$ implies $\beta_X^{(0)} \beta_{X-k}^{(0)} \sigma_{e_0}^2=\tilde{\beta}_X^{(0)} \tilde{\beta}_{X-k}^{(0)} \tilde{\sigma}_{e_0}^2$ for $k=1,\ldots,X$. Together with (\ref{eq proof cohort X and X-1}) we can conclude, as in the proof of theorem \ref{thm:LC_rw}, that
either $\sign(\beta_x^{(0)})= \sign(\tilde{\beta}_x^{(0)})$, $\forall x \in \{0,\ldots,X\}$ or $\sign(\beta_x^{(0)})= -\sign(\tilde{\beta}_x^{(0)})$, $\forall x \in \{0,\ldots,X\}$ whenever $g_{\boldsymbol{\theta}}(x,y,s,t)=g_{\tilde{\boldsymbol{\theta}}}(x,y,s,t)$. We then obtain as in the proof of theorem \ref{thm:LC_rw} that $\mathrm{sgn}(\tilde{\beta}_{x}^{(0)}) = \mathrm{sgn}(\beta_{x}^{(0)})$ and $c=1$, which implies $\tilde{\beta}_{x}^{(0)} = \beta_{x}^{(0)}$ and $\tilde{\sigma}_{e_0}^2 = \sigma_{e_0}^2$. In view of expression (\ref{eq covariance cohort plug-in}) this result together with $g_{\boldsymbol{\theta}}(x,y,s,t)=g_{\tilde{\boldsymbol{\theta}}}(x,y,s,t)$ 
entails the restriction
\begin{equation}
	\label{eq:LCcohortrestr2}
	\tilde{\beta}_x^{(1)} \tilde{\beta}_y^{(1)} \tilde{\sigma}_{e_1}^2 = \beta_x^{(1)} \beta_y^{(1)} \sigma_{e_1}^2, \, \forall x,y=0,\ldots, X.
\end{equation}
Analogously to the case of $\tilde{\beta}_{x}^{(0)}$ and $\tilde{\sigma}_{e_0}^2$, only alternative parameterisations that satisfy $|\tilde{\beta}_x^{(1)}|=c|\beta_x^{(1)}|$, $x=0,\ldots,X$ are compatible with \eqref{eq:LCcohortrestr2}. The same reasoning as in the case of $\tilde{\beta}_{x}^{(0)}$ and $\tilde{\sigma}_{e_0}^2$ shows $\tilde{\beta}_{x}^{(1)} = \beta_{x}^{(1)}$ and $\tilde{\sigma}_{e_1}^2 = \sigma_{e_1}^2$. Hence, $( \tilde{{\beta}}^{(0)}, \tilde{{\beta}}^{(1)}, \tilde{\sigma}_{e_0}^2,  \tilde{\sigma}_{e_1}^2,  \tilde{\sigma}_{\epsilon}^2) = ({\beta}^{(0)},{\beta}^{(0)}, \sigma_{e_0}^2,  \sigma_{e_1}^2,  \sigma_{\epsilon}^2)$.

Consider now the expected values. Since $(\tilde{\beta}^{(0)},\tilde{\beta}^{(1)}) = (\beta^{(0)},\beta^{(1)})$ equation (\ref{eq expected values cohort random walk}) boils down to
\begin{equation}\label{eq:noidentcond}
			\alpha_x - \tilde{\alpha}_x + \beta_x^{(0)}(\mu_0 - \tilde{\mu}_0)(X-x) = -(\beta_x^{(0)}(\mu_0 - \tilde{\mu}_0) + \beta_x^{(1)}(\mu_1 - \tilde{\mu}_1))t.
\end{equation}
Because the left-hand side does not depend on $t$ and $T \geq 4$, both sides of the above equation must equal zero.
 Hence, 
  $\beta_x^{(0)}(\mu_0 - \tilde{\mu}_0) + \beta_x^{(1)}(\mu_1 - \tilde{\mu}_1)=0$. If there is a $\bar{x} \in \{0,\ldots,X\}$ such that $\beta_{\bar{x}}^{(0)}=0$ but $\beta_{\bar{x}}^{(1)} \neq 0$ or such that $\beta_{\bar{x}}^{(1)}=0$ but $\beta_{\bar{x}}^{(0)} \neq 0$ we obtain $\mu_1=\tilde{\mu}_1$ or $\mu_0=\tilde{\mu}_0$, respectively. Indeed if $\beta_{\bar{x}}^{(0)}=0$ we must have $\mu_1=\tilde{\mu}_1$, because $\beta_{\bar{x}}^{(1)} \neq 0$. Under the constraints there must be at least one $\beta_x^{(0)} \neq 0$ which then implies $\mu_0=\tilde{\mu}_0$. Similar if $\beta_{\bar{x}}^{(1)}=0$ but $\beta_{\bar{x}}^{(0)} \neq 0$. Hence, it remains to consider that $\forall x \in \{0,\ldots,X\}$ we have either $\beta_x^{(0)}=0 \wedge \beta_x^{(1)}=0$ or $\beta_x^{(0)} \neq 0 \wedge \beta_x^{(1)} \neq 0$. Whenever $\beta_x^{(0)} \neq 0$ we can rewrite $
	\beta_x^{(0)}(\mu_0 - \tilde{\mu}_0) + \beta_x^{(1)}(\mu_1 - \tilde{\mu}_1)=0
$
as
$
	\tilde{\mu}_0 = \mu_0 + (\beta_x^{(1)} \slash \beta_x^{(0)})(\mu_1 - \tilde{\mu}_1).
$	
Here $\tilde{\mu}_0$ does not depend on $x$, implying that $\beta_x^{(1)} \slash \beta_x^{(0)}$ is constant as a function of $x$, i.e.~$\beta_x^{(1)} = d \beta_x^{(0)}$, $\forall x$, for some constant $d$. However only $d=1$ satisfies the identifying restrictions made in the model, i.e.~$\beta_x^{(1)} = \beta_x^{(0)},\,\forall x$. But there is no $\boldsymbol{\theta} \in \Theta_{(0,1,0) \times (0,1,0)}^{ct}$ such that $\beta^{(0)} = \beta^{(1)}$, which now implies $\tilde{\mu}_1 = \mu_1$ and $\tilde{\mu}_0 = \mu_0$.
Given these latter results, the left-hand side of \eqref{eq:noidentcond} reduces to $\alpha_x - \tilde{\alpha}_x = 0$ whose only solution is $\alpha_x = \tilde{\alpha}_x$.
\end{proof}

\subsection{ARIMA(1,1,0) plug-in Lee Carter model and proof of theorem \ref{prop:LC_ARIMA110}}\label{appendix ARIMA(1,1,0)}
Before presenting the proof of theorem \ref{prop:LC_ARIMA110} we give some background information on the autoregressive integrated moving average process with parameter (1,1,0). If $(y_t)$ is such a process we have with $L$ being the lag-operator  $(1- \rho L) \Delta y_t - \mu = e_t$ which can be written as
\begin{equation}\label{eq appendix arima(1,1,0)}
	y_t = y_0 + \mu t + \sum_{s=1}^t \sum_{\ell = 0}^\infty \rho^{\ell} e_{s-\ell};
\end{equation}
see, for instance, \citep[Section 9.1]{BrockwellDavis2006}.\\
For the Lee-Carter model
\begin{align*}
	\log(m_{x,t}) &= \alpha_x + \beta_x \kappa_t + \epsilon_{x,t}, \quad x = 1,\ldots, X, \, t = 1, \ldots, T,
\end{align*}
we obtain upon replacing $(\kappa_t)$ by an autoregressive integrated moving average process with parameter (1,1,0)
\begin{equation*} 
	\log(m_{x,t}) = \alpha_x + \beta_x c + \beta_x\mu t + \beta_x \sum_{s=1}^t \sum_{\ell = 0}^\infty \rho_1^{\ell} e_{s-\ell} + \epsilon_{x,t},
\end{equation*}
where $c=\kappa_0$. This is equation (\ref{eq:LCARIMA110}) in section \ref{section extensions}. The first moments and the covariance structure of this model are directly obtained from the below  derivations for $(y_t)$.\\

The expected value of the process defined by (\ref{eq appendix arima(1,1,0)}) is simply given by $\E(y_t) = y_0 + \mu t$. Let $t\wedge q = \min\{t,q\}$ and $t\vee q = \max\{t,q\}$. Then, for the second moments of this process we find
\begin{align*}
	\covi(y_t, y_q) &= \E\left[\left(\sum_{s=1}^t \sum_{\ell = 0}^\infty \rho^{\ell} \, e_{s-\ell}\right)\left(\sum_{r=1}^q \sum_{k = 0}^\infty \rho^{k} \,e_{r-k}\right)\right] \\
	&= \sum_{s=1}^t \sum_{r=1}^q \sum_{\ell = 0}^\infty \sum_{k = 0}^\infty \rho^{k+\ell} \E(e_{s-\ell}e_{r-k}) \\
	&= \sum_{s=1}^{t\wedge q} \sum_{\ell = 0}^\infty \sum_{k = 0}^\infty \rho^{k+\ell} \E(e_{s-\ell}e_{s-k}) + 2\sum_{s=2}^{t\wedge q} \sum_{r=1}^{s-1} \sum_{\ell = 0}^\infty \sum_{k = 0}^\infty \rho^{k+\ell} \E(e_{s-\ell}e_{r-k}) \\
	& \hspace*{0.4cm} + \sum_{s=(t\wedge q)+1}^{t\vee q} \sum_{r=1}^{t\wedge q} \sum_{\ell = 0}^\infty \sum_{k = 0}^\infty \rho^{k+\ell} \E(e_{s-\ell}e_{r-k}) \\
	& = I+II+III.
\end{align*}
Using Hamilton(1994, Equations (3.4.4) and (3.4.5)) for the covariances of an autoregressive process with lag 1, we obtain
		\begin{align*}
		I & = (t\wedge q) \, \frac{\sigma^2}{1-\rho^2}
		\intertext{and}
		II &= 2\frac{\sigma^2}{1-\rho^2} \sum_{s=2}^t \sum_{r=1}^{s-1} \rho^{s-r},
\end{align*}
{Using the geometric series formula twice} the sum of these two expression can be simplified to yield
\begin{align*}
	I+II & = (t \wedge q) \, \frac{\sigma^2}{1-\rho^2} + 2\frac{\sigma^2}{1-\rho^2} \left(  \frac{\rho^{t\wedge q+1} - \rho}{(\rho -1)^2} + \frac{(t\wedge q) \rho (1-\rho)}{(\rho-1)^2} \right) \\
	& = \sigma^2 \frac{(\rho^2-1)(t\wedge q) - 2\rho(\rho^{t\wedge q} -1)}{(\rho-1)^3(\rho+1)} \\
	& = (t\wedge q) \, \frac{\sigma^2}{(1-\rho)^2} - 2\sigma^2 \rho \frac{(\rho^{t\wedge q}-1)}{(\rho-1)^3(\rho+1)}.
\end{align*}
With regards to $III$, we can apply the geometric series formula again to obtain
\begin{equation*}
	III = \sigma^2\rho \frac{\rho^{t\wedge q} - \rho^{t\vee q} + \rho^{|t-q|} - 1}{(\rho-1)^3(\rho+1)}.
\end{equation*}
Plugging this result into the main expression of the covariance and re-arranging, we obtain
\begin{equation}\label{eq cov arima(1,1,0)}
	\covi(y_t, y_q) = (t \wedge q)\frac{\sigma^2}{(1-\rho)^2} -  \sigma^2 \frac{\rho(\rho^{t \vee q} - \rho^{|t-q|} +  \rho^{t \wedge q}-1)}{(\rho-1)^3(\rho+1)}.
\end{equation}

We now turn to the proof of theorem \ref{prop:LC_ARIMA110}.\\

\bigskip

\begin{proof}{\bf of theorem \ref{prop:LC_ARIMA110}} We start by looking at the variances of \eqref{eq:LCARIMA110}. Combining it with I and II from the above we have
\begin{align*} 
		\vari(\log(m_{x,t})) & = \beta_x^2 \sum_{\ell=1}^t \sum_{s=1}^t \sum_{k=0}^{\infty} \sum_{r=0}^{\infty} \rho^{k+r}\E(e_{\ell-k} e_{s-r}) + \sigma_\epsilon^2 \notag \\
		&= \beta_x^2 \left(t \frac{\sigma_e^2}{1-\rho^2} + 2 \frac{\sigma_e^2}{1-\rho^2} \sum_{\ell=2}^t \sum_{k=1}^{\ell-1} \rho^{\ell-k}\right) + \sigma_\epsilon^2,
\end{align*}
which for fixed $x$ is the sum of a constant and two terms that depend on $t$. Consequently, if we have a parameter $\tilde{\boldsymbol{\theta}}$ that leads to the same covariance structure as $\boldsymbol{\theta}$ we can write
	\begin{align}\label{eq:LCARIMArestr01}
		\beta_x^2 \left(t \frac{\sigma_e^2}{1-\rho^2} + 2 \frac{\sigma_e^2}{1-\rho^2} \sum_{\ell=2}^t \sum_{k=1}^{\ell-1} \rho^{\ell-k}\right) = \tilde{\beta}_x^2 \left(t \frac{\tilde{\sigma}_e^2}{1-\tilde{\rho}^2} + 2 \frac{\tilde{\sigma}_e^2}{1-\tilde{\rho}^2} \sum_{\ell=2}^t \sum_{k=1}^{\ell-1} \tilde{\rho}^{\ell-k}\right) + z
		\intertext{or}
		\label{eq:LCARIMArestr0}
		t \left(\beta_x^2 \frac{\sigma_e^2}{1-\rho^2} - \tilde{\beta}_x^2 \frac{\tilde{\sigma}_e^2}{1-\tilde{\rho}^2}\right) = -2\left(\beta_x^2\frac{\sigma_e^2}{1-\rho^2} \sum_{\ell=2}^t \sum_{k=1}^{\ell-1} \rho^{\ell-k} - \tilde{\beta}_x^2\frac{\tilde{\sigma}_e^2}{1-\tilde{\rho}^2} \sum_{\ell=2}^t \sum_{k=1}^{\ell-1} \tilde{\rho}^{\ell-k}\right) + z
	\end{align}
with $z=\tilde{\sigma}_\epsilon^2 - \sigma_\epsilon^2$. Because the left-hand side of (\ref{eq:LCARIMArestr0}) is an affine linear function in $t$
taking the second differences of both sides yields
	\begin{equation}\label{eq exponential functions}
		 \frac{\beta_x^{2}\sigma_e^2}{1-\rho}\rho^{t-1} - \frac{\tilde{\beta}_x^{2}\tilde{\sigma}_e^2}{1-\tilde{\rho}}\tilde{\rho}^{t-1} = 0,\,t=3,\ldots,T
	\end{equation}	
which, unless $\beta_x=\tilde{\beta}_x=0$ or $\rho=\tilde{\rho}=0$, can only be zero if $\tilde{\rho}= \rho$ and $\tilde{\beta}_x^2\tilde{\sigma}_e^2 = \beta_x^2\sigma_e^2$ as $T \geq 4$. If $\beta_x=\tilde{\beta}_x=0$ for some $x$ we know that under the constraint $\sum_{x=0}^X \beta_x=1$ there must be at least one $\bar{x} \in \{0,\ldots,X\}$ such that $\beta_{\bar{x}} \neq 0$. From equation (\ref{eq exponential functions}) with $x=\bar{x}$ we then obtain $\rho=\tilde{\rho}$. If $\rho=\tilde{\rho}=0$ we have the model considered in theorem \ref{thm:LC_rw} and we can proceed as we did there  to show that $\boldsymbol{\theta}=\tilde{\boldsymbol{\theta}}$. To conclude equation (\ref{eq exponential functions}) implies $\tilde{\rho} = \rho$ and $\tilde{\beta}_x^2\tilde{\sigma}_e^2 = \beta_x^2\sigma_e^2,\, x=0,\ldots,X$, and we can restrict ourselves to the case $\rho=\tilde{\rho}\neq 0$.\\
 As in the proof of theorem \ref{thm:LC_rw} identifiability now follows directly for $X=0$. For $X \geq 1$, we only need to consider $x\in \breve{X}$ with $\breve{X}$ as defined in the aforementioned proof. Consider now $\covi(\log(m_{x,s}), \log(m_{y,T}))$ for $x \neq y$. Combining (\ref{eq:LCARIMA110}) and (\ref{eq cov arima(1,1,0)}) 
 \begin{align}\label{eq:LCARIMA110cov}
 			 \covi(\log(m_{x,s}), \log(m_{y,T})) &= \sigma_e^2\beta_x \beta_y \left(\frac{s}{(1- \rho)^2} - \frac{\rho(\rho^T + \rho^s - \rho^{T-s}-1)}{(\rho-1)^3(\rho+1)}\right) \nonumber \\
 			& x, y \in \breve{X}, \, x \neq y, \,s \in \{1,\ldots,T\}.
 \end{align}
 First, we need to make sure that there is no $\rho$, $\rho \neq 0$, such that the term in parentheses is zero for all $s=1,\dots, T-1$.
 If this is the case the first differences of the expression with respect to $s$ are zero as well, i.e.
 \begin{equation*}
 			\frac{1}{(1-\rho)^2} - {\frac{\rho(\rho^s(1-\rho^{-1})+\rho^{T-s}(\rho-1))}{(\rho-1)^3(\rho+1)}} = 0.
 \end{equation*}
 Solving for the terms that involve $s$ yields
 \begin{equation*}
 \rho^s(1-\rho^{-1})+\rho^{T-s}(\rho-1) = \rho^{-1}((1-\rho)^{-2})(\rho-1)^3(\rho+1),
  \end{equation*}
 where the exact expression on the right-hand side can be disregarded. Most important is that the left-hand side must be constant for all $s=1,\ldots,T-1$. To check this restriction, we simply set equal the cases $s=1$ and $T-1$, yielding
 \begin{align*}
 	\rho(1-\rho^{-1})+\rho^{T-1}(\rho-1) &= \rho^{T-1}(1-\rho^{-1})+\rho(\rho-1) \\
 	\Leftrightarrow 1 &=\rho^{T-2}.
 \end{align*}
Obviously, for $T>2$ the only solutions to this equality are $\rho=0$ and $\rho=1$ as well as $\rho=-1$ for odd $T$. However, none of these values are elements of the parameter space defined below equation \eqref{eq:LCARIMA110}. Hence, for $T > 2$ there are no permissible values for $\rho$ such that the term in parentheses in equation \eqref{eq:LCARIMA110cov}
equals zero.
		
Given the knowledge that the covariances are non-zero (if $x,y \in \breve{X}$) and that $\tilde{\rho} = \rho$, a valid alternative parameterisation of the second moments must hence satisfy $\sigma_e^2\beta_x \beta_y = \tilde{\sigma}_e^2\tilde{\beta}_x \tilde{\beta_y}$. This restriction is identical to that found in the proof of theorem \ref{thm:LC_rw} and it can be proceeded in exactly the same way as done there to obtain $\tilde{\boldsymbol{\beta}} = \boldsymbol{\beta}$ and $\tilde{\sigma}_e^2 = \sigma_e^2$. We finish the proof by reconsidering \eqref{eq:LCARIMArestr01}. Given the results established in this proof, the equality holds only if $z = 0$ which implies $\tilde{\sigma}_{\epsilon}^2 = \sigma_\epsilon^2$.

It remains to consider the expected values. From \eqref{eq:LCARIMA110}, we have
\begin{align*}
	\E(\log(m_{x,t})) = \alpha_x + \beta_x c + \beta_x \mu t.
\end{align*}
This expression is identical to equation (\ref{eq identifiability proof means}), which contained the expected values of the ARIMA(0,1,0) model. Together with $\tilde{\boldsymbol{\beta}} = \boldsymbol{\beta}$, this provides the same starting point for showing that $\tilde{\boldsymbol{\alpha}} = \boldsymbol{\alpha}$ and $\tilde{\mu} = \mu$ as done in the proof of theorem \ref{thm:LC_rw}.
\end{proof}

\bigskip

\subsection{ARIMA(0,1,1) plug-in Lee Carter model and proof of theorem \ref{prop:LC_ARIMA011}}\label{appendix ARIMA(0,1,1)}

The ARIMA(0,1,1) model is given by $\Delta y_{t}-\mu=(1+\phi_{1}L)e_{t}$
so that the process in levels is expressed by $$y_{t}=y_{0}+\mu\,t+\sum_{s=1}^{t}(e_{s}+\phi_{1}e_{s-1}).$$

Given that the expected value of this process is $\E(y_{t})=y_{0}+\mu t$,
the covariances are given by
\begin{align*}
\covi(y_{t},y_{q}) & =\E(\sum_{s=1}^{t}\sum_{r=1}^{q}(e_{s}+\phi_{1}e_{s-1})(e_{r}+\phi_{1}e_{r-1})
\end{align*}

Again, let $t\vee q=\max\{t,q\}$ and $t\wedge q=\min\{t,q\}$. Furthermore, define $A(s,r)=e_{s}e_{r}+\phi_{1}e_{s-1}e_{r}+\phi_{1}e_{s}e_{r-1}+\phi_{1}^{2}e_{s-1}e_{r-1}$.
We can decompose the above sum into
\begin{align*}
\covi(y_{t},y_{q}) & =\sum_{s=1}^{t\wedge q}\E(A(s,s))+\sum_{s=1}^{t\wedge q}\sum_{r\ne s}^{t\wedge q}\E(A(s,r))+\sum_{s=(t\wedge q)+1}^{t\vee q}\sum_{r=1}^{t\wedge q}\E(A(s,r))\\
 & =I+II+III.
\end{align*}

For $I$ we directly obtain $I=\sigma_{e}^{2}(t\wedge q)(1+\phi_{1}^{2})$.
$III$ reduces to
\begin{align*}
III & =\eins_{\{t\ne q\}}(t,q)\sigma_{e}^{2}\phi_{1}
\end{align*}
because $\E(A((t\wedge q)+1,t\wedge q))$ is the only nonzero
term involved. Lastly, $II=2((t\wedge q)-1)\sigma_{e}^{2}\phi_{1}$
since only $\E(A(s,s-1))$ and $\E(A(r-1,r))$ are
nonzero. Consequently, $\covi(y_{t},y_{q})=\sigma_{e}^{2}(t\wedge q)(1+\phi_{1}^{2})+2((t\wedge q)-1)\sigma_{e}^{2}\phi_{1}+\eins_{\{t\neq q\}}(t,q)\sigma_{e}^{2}\phi_{1}$
which can be expressed alternatively as
\begin{align*}
\covi(y_{t},y_{q}) & =
\sigma_{e}^{2} \left[ (t\wedge q) (\phi_1 +1)^2 - (1 + \eins_{\{t=q\}}(t,q))\phi_1 \right].
\end{align*}

Using this expression within the Lee-Carter model yields
\begin{align}
\label{eq:LCcovs}
&\covi(\log(m_{x,q}),\log(m_{y,t})) \notag \\  &= 
\beta_{x}\beta_{y} \sigma_{e}^{2} \left[ (t\wedge q) (\phi_1 +1)^2 - (1 + \eins_{\{t=q\}}(t,q))\phi_1 \right] + \eins_{\{t=q\}}(t,q) \eins_{\{x=y\}}(x,y)\sigma_{\epsilon}^{2}.
\end{align}

This result allows us to turn to the proof of theorem \ref{prop:LC_ARIMA011}.

\bigskip

\begin{proof}{\bf of theorem \ref{prop:LC_ARIMA011}}
Consider the covariance structure \eqref{eq:LCcovs}.
For $s=t$ and fixed $x,y$ such that $x\ne y$ the above expression reduces to
\begin{equation*}
	\covi(\log(m_{x,t}),\log(m_{y,t}))=\beta_x \beta_y \sigma^2_e  \left[ t (\phi_1 +1)^2 - 2\phi_1 \right],
\end{equation*}
which is an affine function in $t$. Since this function can be scaled arbitrarily by a factor $z$ that is compensated by $\tilde{\beta}_x \tilde{\beta}_y\tilde{\sigma}_e^2$, we have the restrictions
\begin{align*}
z(1+\tilde{\phi}_1)^{2} & =(1+\phi_1)^{2}\\
z\tilde{\phi}_1 & =\phi_1 ,
\end{align*}
which yield
$$
\frac{(1+\phi_1)^{2}}{(1+\tilde{\phi}_1)^{2}}=\frac{\phi_1}{\tilde{\phi}_1}.
$$
Solving this equation in $\tilde{\phi}_1$ for a given $\phi_1$ we find that either $\tilde{\phi}_1=\phi_1$ or $\tilde{\phi}_1= 1 \slash \phi_1$ for $\phi_1 \neq 0$. However, the second solution does not belong to $\Theta_{(0,1,1)}$. Consequently, $\tilde{\phi}_1 = \phi_1$. The remainder of this proof is identical to the proof of theorem \ref{thm:LC_rw} as the problem now reduces to $\beta_x \beta_y \sigma^2_e=\tilde{\beta}_x \tilde{\beta}_y \tilde{\sigma}^2_e$.
\end{proof}



\end{document}